\documentclass[twocolumn,floatfix,preprintnumbers,aps,prd,showpacs,nofootinbib,amsmath,amssymb,amsfonts]{revtex4}
\usepackage{bm}

\usepackage{graphicx}
\usepackage{color}
\usepackage[pdfborder={0 0 0},colorlinks=false]{hyperref}
\usepackage{dsfont}
\usepackage{here}

\newcommand{\etal}{{\em et al.}}

\newcommand{\be}{\begin{equation}}
\newcommand{\ee}{\end{equation}}
\newcommand{\bea}{\begin{eqnarray}}
\newcommand{\eea}{\end{eqnarray}}
\newcommand{\gr}[1]{\bm{#1}}
\newcommand{\ii}{\mathrm{i}}
\newcommand{\dd}{\mathrm{d}}

\begin{document}
\title{Weak lensing $B$-modes on all scales as a probe of local isotropy}

\author{Cyril Pitrou}

\author{Jean-Philippe Uzan}
 \affiliation{
             Institut d'Astrophysique de Paris,
             Universit\'e Pierre~\&~Marie Curie - Paris VI,
             CNRS-UMR 7095, 98 bis, Bd Arago, 75014 Paris, France,}

\author{Thiago S. Pereira}
\affiliation{Departamento de F\'isica, Universidade Estadual de Londrina,
 86051-990, Londrina, Paran\'a, Brazil.}

\date{\today}

\begin{abstract}
This article introduces a new multipolar hierarchy for the propagation of the weak-lensing shear, convergence, and twist valid in a general spacetime. Our approach is fully covariant and relies on no perturbative expansion. We show that the origin of $B$-modes, in particular on large angular scales, is related to deviations of isotropy of the spacetime. Known results assuming a Friedmann-Lema\^{\i}tre background spacetime are naturally recovered. The example of a Bianchi~$I$ spacetime illustrates our formalism and its implications for future observations are stressed.
\end{abstract}
\maketitle

\section{Introduction}

Weak gravitational lensing by the large-scale structure of the
Universe has now become a major tool of cosmology~\cite{revuesWL},
used to study questions ranging from the distribution of dark matter
to tests of general relativity~\cite{testGR}. The standard
lore~\cite{refbooks,stebbins} states that, in a homogeneous and
isotropic spacetime, weak lensing effects induce a shear field which,
to leading order, only contains $E$-modes so that the measured level
of $B$-modes is used as an important sanity check at the end of the
data processing chain. $B$-modes contribution to the observed shear can be related to intrinsic alignments~\cite{spinG}, Born correction and lens-lens coupling~\cite{lenslenssimu,cooray}, and gravitational lensing due to the redshift clustering of source galaxies~\cite{clustering}. From an observational point of view, the separation of $E$- and $B$-modes requires in principle to measure the shear correlation at zero separation~\cite{EBmix,EBseparation} that can be brought down to the percent-level accuracy, e.g. with CFHTLenS data~\cite{obs1}. 

This paper emphasizes that any deviation from local spatial isotropy, as assumed in the standard cosmological framework in which the background spacetime is described by a Friedmann-Lema\^{\i}tre (FL) universe, induces $B$-modes in the shear field. More importantly, and contrary to the above mentioned effects, these $B$-modes arise on 
{\it all} cosmological scales. Therefore, any bound on their level can be used as a constraint on spatial isotropy. This is an important signature which, in principle, can be exploited in order to disentangle this geometrical origin of $B$-modes from other non-cosmological effects~\cite{next}. Since it is important for future surveys to predict the level at which these cosmological effects produce $B$-modes, we introduce in this work a new multipolar hierarchy for the weak-lensing shear, convergence and twist that does not assume a specific background geometry. This approach will allow us to pinpoint the origin of the $B$-modes and, in a future work, to access the magnitude of currently observed level of $B$-modes.

This work is organized as follows: we start in \S~\ref{GeodesicBundle} by reviewing the basic formalism of weak-gravitational lensing, which will also help us to set up the basic notations and conventions. In \S\ref{shear-convergence} we derive the evolution equations for the irreducible components of the Jacobi map, which are then used to derive the main multipole expansion hierarchy in \S~\ref{multipole}. We then show how the standard FL results are recovered (\S~\ref{FL}) and discuss the particular case of a Bianchi $I$  (B$I$) universe (\S~\ref{BI}). Finally, we present our conclusion in \S~\ref{conclusion}.\\

Throughout this paper we work with units in which $c=\hbar=1$. Spacetime indices are represented by Greek letters. Upper case Latin indices such as $\{I,J,K,\dots\}$ vary from 1 to 3 and represent spatial coordinates. Furthermore components of vectors on a spatial triad (a set of three orthogonal spatial vectors which are normalized to unity) are denoted with lower case Latin indices $\{i,j,k,\dots\}$, whereas the screen projected (two-dimensional) components are represented by indices $\{a,b,c,\dots\}$ which vary from 1 to 2.

\section{Multipolar hierarchy for weak-lensing} \label{GeodesicBundle}

\subsection{Description of the geodesic bundle}\label{sub2A}

\begin{figure}[h]
\includegraphics[width=0.9\columnwidth]{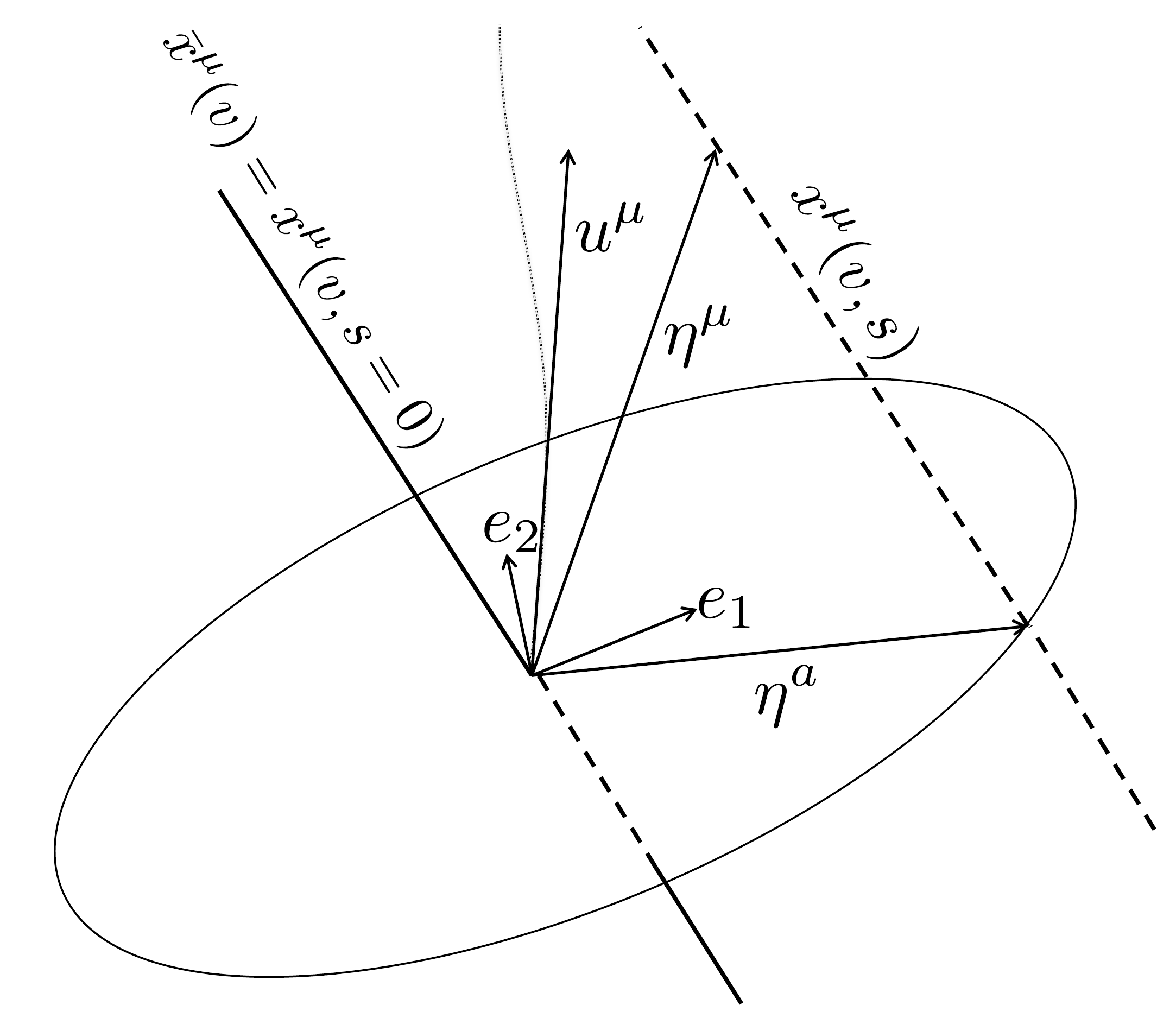}
\caption{Representation of two null geodesics of the light bundle. $\eta^a$ is the projection of $\eta^\mu$ in the plane 
spanned by the basis $\{e_a\}$. The dotted curve represents the worldline of the observer comoving with $u^\mu$. The geodesic bundle is thin so that its transverse dimension has not been depicted and it converges at the observer.}
\label{f0}
\end{figure}

A crucial quantity for weak-lensing is the electromagnetic wave-vector, $k_\mu=\partial_\mu w$, where $w$ is the
phase of the wave. In the eikonal approximation, $k^\mu$ is a null
vector ($k^\mu k_\mu=0$) satisfying a geodesic equation
($k^\nu\nabla_\nu k^\mu=0$). Moreover, if we assume that $\nabla_\mu\nabla_\nu
w=\nabla_\nu\nabla_\mu w$ for any scalar function $w$ (torsion-free hypothesis), it follows that
its integral curves $x^\mu(v)$ defined by $k^\mu(v) \equiv \dd
x^\mu/\dd v$, where $v$ is the affine parameter along
a given geodesic, are irrotational ($\nabla_{[\mu}k_{\nu]}=0$).
Second, we consider a family of null (light-like) geodesics collectively
characterized by $x^\mu(v,s)$, where $s$ labels each member of the family. We adopt
the convention according to which $v=0$ at the observer and increases
toward the source. There is a wave-vector for each geodesic, that is $k^\mu(v,s) \equiv \partial x^\mu /\partial v$, 
and the separation between the geodesics is encompassed by the vector
$\eta^\mu\equiv \partial x^\mu/\partial s$ connecting two neighbor
geodesics (see Fig.~(\ref{f0})). Hence, we first derive the dynamics for a reference geodesic,
and then the dynamics for the deviation vector. 

We suppose that the light-rays converge to a fundamental observer
comoving with the four-velocity $u^\mu$ of matter, which is normalized such that 
$u^\mu u_\mu=-1$. This observer measures a redshift 
$z$ given 
by
\begin{equation}\label{def_z}
  1+z(v)\equiv \frac{(k_\mu u^\mu)_v}{(k_\mu u^\mu)_0}
\end{equation}
so that the energy of the incoming photon is 
\be
U=U_0(1+z)\,,\qquad U_0=(k^\mu u_\mu)_0\,.
\ee
In this work we adopt the perspective of a photon going to the past, which means 
that in a local Lorentz frame, where $u^\mu=(-1,0,0,0)$, we have $k^0=\dd t/\dd v=-U$.
Incidentally, this suggests that we introduce of a ``reduced wave-vector'' through 
\be
\label{khat}
\hat{k}^\mu = \frac{\dd x^\mu}{\dd\hat{v}}\equiv U^{-1} k^\mu
\ee
in order to simplify our expressions \footnote{Since $\dd\hat{v}=-\dd t$, the new parameter $\hat{v}$ 
is simply the negative of the proper time $t$, reflecting our choice of perspective in which the observer sheds light on the source.}

At each position $x^\mu$ of a given geodesic we can associate a direction vector
$\gr{n}$ whose components are $n^\mu$, and defined from the reduced wave-vector through
\begin{equation}
\hat{k}^\mu = -u^\mu + n^\mu\,,
\end{equation}
with
\be
u^\mu n_\mu=0\,,\quad n_\mu n^\mu =1\,.
\ee
At the observer, $\gr{n}^o\equiv\gr{n}(v=0)$ is the spacelike vector
pointing along the line of sight\footnote{Note that our definition of  $n^\mu$ differ by a minus sign 
from that of Ref. \cite{clarkson}}. However, since we now have
$$
\frac{\dd \hat v}{\dd v}=U,
$$
it follows that we can either choose $(\gr{n}^o,v)$ or $(\gr{n}^o,\hat v)$ as independent set of variables to parameterize the geodesic, which correspond to two slices of the past lightcone. As we shall see below, the use of $\hat v$ simplifies the derivation of the multipolar expansion for the weak-lensing observables. 

At a given point of the geodesic, it is necessary to add two vectors
to ${\bm u}$ and ${\bm n}$ in order to obtain a complete basis of the
tangent space. We choose these two vectors $\gr{n}_a$, with $a=\{1,2\}$, to be orthonormalized and
orthogonal to  ${\bm u}$ and ${\bm n}$, that is they are defined by 
\be
\label{basis2d}
n_{a}^\mu n_{b\mu}=\delta_{ab}\,,\quad n_{a}^\mu u_{\mu}=n_{a}^\mu n_{\mu}=0\,,\quad(a=1,2)\,. 
\ee
Since $\gr{n}$ and $\gr{n}_a$ comprise a three-dimensional orthonormal basis, we can simplify the notation by defining $\gr{n}_3\equiv \gr{n}$ so that we can collectively write $\gr{n}_i\equiv \lbrace n^\mu_i\rbrace_{i=1\ldots3}$. Note that at the observer we can again define $\gr{n}_i^o\equiv \gr{n}_i(\hat v=0)$ with a remaining rotation freedom around $\gr{n}^o$ for the choice of $\gr{n}^o_a$.

\begin{figure*}[!htb]
\begin{center}
\includegraphics[width=2.0\columnwidth]{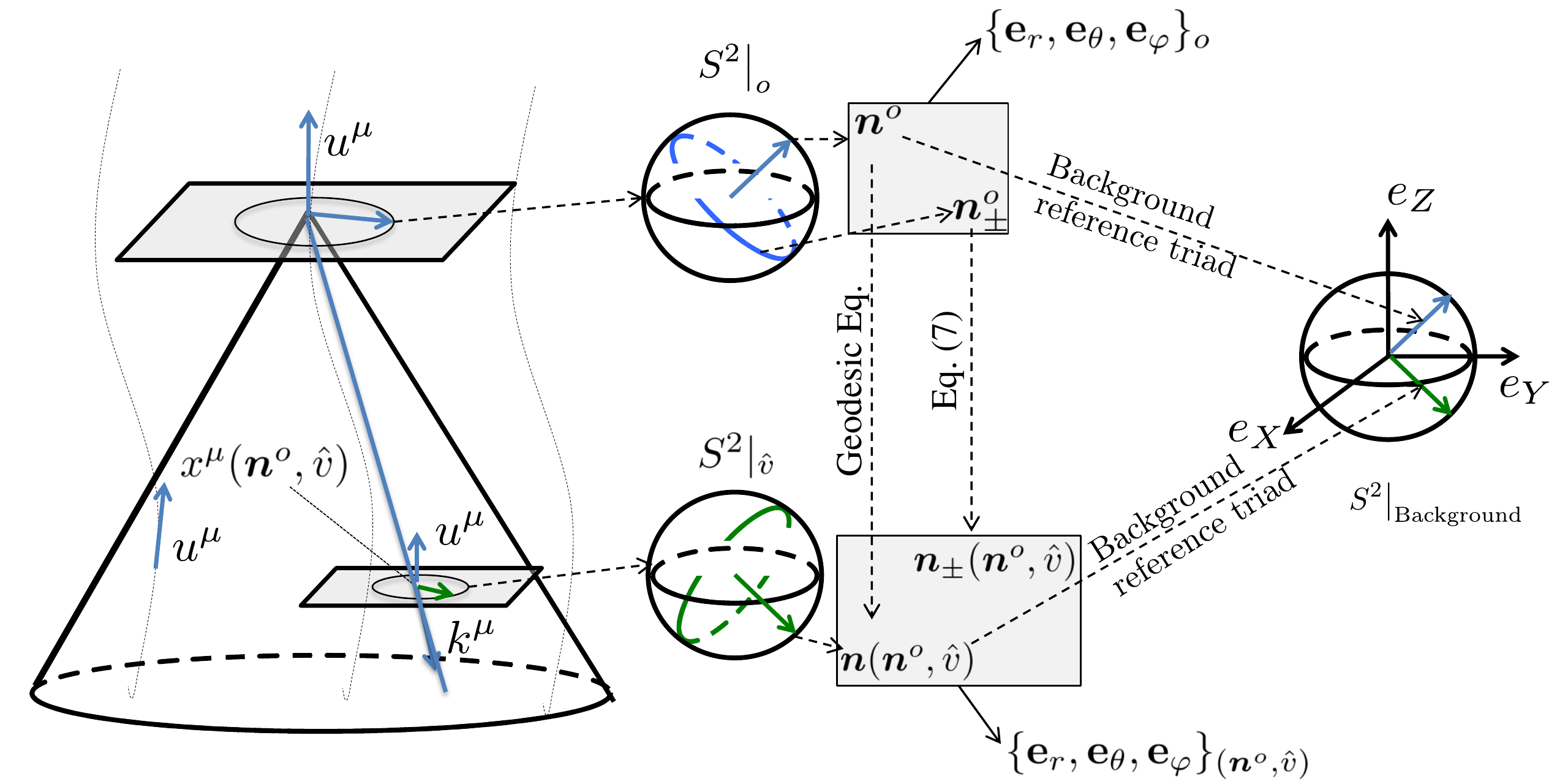}
\caption{Any position on the past lightcone can be considered as $x^\mu(\gr{n}^o,\hat v)$. While quantities such as ${\cal E}_{\mu\nu}$ are local, quantities such as ${\cal W}_{ab}$ depend on the local basis at $x^\mu(\gr{n}^o,\hat v)$ via the projection on $n_{\langle a}^\mu n_{b\rangle}^\nu$; see Eq.~(\ref{proj_W}). Observational quantities are however defined in terms of $\gr{n}^o$ so that one needs to relate the basis $\lbrace\gr{e}_r,\gr{e}_\theta,\gr{e}_\varphi\rbrace$ in $(\gr{n}^o,\hat v)$ and in $\hat v=0$. The relation $\gr{n}(\gr{n}^o,\hat v)$ induces ``projection effects'' and a non-local relation between quantities like  ${\cal E}_{\ell m}$ and ${\cal W}_{ab}$. Once a background spacetime is chosen, its symmetries simplify the comparison. For instance, a B$I$ spacetime provides a natural triad of Killing vectors associated to its principal axis. One can use this ``global reference'' to relate the local $S^2$ in $x^\mu(\gr{n}^o,\hat v)$ to the observer's $S^2$ by comparing them in the reference $S^2$.}
\label{f1}
\end{center}
\end{figure*}

We now introduce the screen projector tensor
\begin{equation}
  S_{\mu\nu}\equiv g_{\mu\nu}+u_\mu u_\nu -n_\mu n_\nu, 
\end{equation}
which projects any tensor on the two-dimensional surface orthogonal to the line of sight. Thanks to the orthogonality relations (\ref{basis2d}),  the basis can be parallel transported along the null geodesic as~\cite{Lewis2006} 
\begin{equation}\label{e:prop_n}
  S_{\mu \sigma} k^\nu \nabla_\nu n_a^\sigma=0.
\end{equation}
At $\hat v=0$, each $\gr{n}^o$ of the geodesic bundle can be associated to a spherical basis and
this can be used to fix the rotational freedom. Indeed, for each $\gr{n}^o$ there will be a
unique choice of $\gr{n}^o_1(\gr{n}^o)$ and $\gr{n}^o_2(\gr{n}^o)$ if we set $\{\gr{e}^o_r,\gr{e}^o_\theta,\gr{e}^o_\varphi\}=\{\gr{n}^o,\gr{n}_1^o,\gr{n}_2^o\}$. The integration of Eq.~(\ref{e:prop_n}) then allows to define this basis at each point on the past lightcone, i.e. to determine $\gr{n}_i(\gr{n}^o,\hat v)$, or, equivalently, $\{\gr{e}_r,\gr{e}_\theta,\gr{e}_\varphi\}(\gr{n}^o,\hat v)$ everywhere. This prescription emphasizes the importance of introducing a reference triad as a way of identifying these projection effects; see Fig.~\ref{f1}. 

At this point it is convenient to introduce the helicity basis defined as
\begin{equation}
\label{e:helicity}
 \gr{e}_\pm =\gr{n}_\pm\equiv \frac{1}{\sqrt{2}}\left(\gr{e}_\theta \mp \ii \gr{e}_\varphi\right)=\frac{1}{\sqrt{2}}\left(\gr{n}_1 \mp \ii \gr{n}_2\right).
\end{equation}
Their components in the $\gr{n}_a$ basis read simply 
\be
n_\pm^a=\gr{n}_\pm.\gr{n}_a=\frac{1}{\sqrt{2}}(\delta_1^a \mp \ii \delta_2^a) 
\ee
and are, by construction, constant.

We now note that any event on the lightcone is uniquely specified by $(\gr{n}^o,\hat v)$, i.e. it is of the form $x^\mu(\gr{n}^o,\hat v)$. This means that any local quantity $X(x^\mu)$ evaluated on the lightcone can be seen as a function $X(\gr{n}^o,\hat v)$. The redshift defined in Eq.~(\ref{def_z}) is also a function of $(\gr{n}^o,\hat v)$, and $U$ propagates as (see e.g. Ref.~\cite{clarkson})
\begin{equation}
\label{eqHparallel}
 \frac{\dd\ln U}{\dd\hat v} = H_\parallel(\gr{n}^o,\hat v) 
\end{equation}
where the parallel Hubble expansion rate along the line of sight is defined by 
\bea
H_\parallel(\gr{n}^o,\hat v)&\equiv&\hat k^\mu \hat k^\nu\nabla_\mu u_\nu\,.
\eea
Using the standard $1+3$ decomposition of $\nabla_\mu u_\nu$, it takes the general form
\bea\label{e.Hperp}
H_\parallel(\gr{n}^o,\hat v) &=& \frac13 \Theta +\hat\sigma _{\mu\nu}n^\mu n^\nu+ A_\mu n^\mu\,,
\eea
where $\Theta$, $\hat\sigma_{\mu\nu}$ and $A^\mu$ are the expansion, shear and acceleration of the flow  $u^\mu$. All these quantities are evaluated on $[x^\mu(\gr{n}^o,\hat v)]$ and are thus functions of $\gr{n}(\gr{n}^o,\hat v)$ on the past lightcone.

\subsection{Shear, twist and convergence propagation}\label{shear-convergence}

The purpose of this section is to derive an equation governing the shear, twist and convergence of a light-ray bundle without specifying the spacetime structure. The evolution of the deviation vector $\eta^\mu$ is given by the
geodesic {\it deviation} equation
\be
\frac{\dd^2\eta^\mu}{\dd s^2} = {R^\mu}_{\nu\alpha\beta} k^\nu k^\alpha\eta^\beta\,,
\ee
where ${R^\mu}_{\nu\alpha\beta}$ is the Riemann tensor. This equation can be rewritten in terms of its component on the screen basis $\lbrace \gr{n}_a\rbrace$ 
as~\cite{refbooks}
\begin{equation}\label{e.gde}
\frac{\dd^2\eta_a}{\dd v^2} = {\cal R}_{ab}\eta^b\,,
\end{equation}
where
\begin{equation}
  {\cal R}_{ab}\equiv{R}_{\mu\nu\alpha\beta}k^\nu k^\alpha n_a^\mu n_b^\beta
\end{equation} 
is the screen projected Riemann tensor. The linearity of Eq.~(\ref{e.gde}) implies that 
$$
  \eta^a(v) = {\cal D}^a_b(v) \left(\frac{{\dd\eta^b}}{{\dd v}}\right)_{v=0}\,,
$$ 
where the Jacobi map ${\cal D}_{Ôab}$ satisfies the Sachs equation~\cite{sachs,refbooks}
\begin{equation}\label{gde2}
   \frac{\dd^2}{\dd v^2}{\cal D}^a_b={\cal R}^a_c{\cal D}^c_b\,,
\end{equation}
subject to the initial conditions
\be
\label{initialconditions}
{\cal D}^a_b(0) =0\,,\quad \frac{\dd{\cal D}^a_b}{\dd v}(0)=\delta^a_b\,.
\ee

In order to proceed, we need to decompose both ${\cal D}_{ab}$ and ${\cal R}_{ab}$ in their irreducible pieces. We start by decomposing the projected Ricci tensor into a trace and a traceless part as 
\be
{\cal R}_{ab}=  U^2\left({\cal R}I_{ab} + {\cal W}_{ab}\right)
\ee
where ${\cal R}$ and ${\cal W}_{ab}$ are related to the Ricci ($R_{\mu\nu}$) and Weyl ($C_{\mu\rho\sigma\nu}$) tensors through:
\be
{\cal R}\equiv - \frac12 R_{\mu\nu}\hat k^\mu \hat k^\nu\,,\quad 
{\cal W}_{ab}\equiv C_{\mu\rho\sigma\nu}\hat k^\rho \hat k^\sigma n_a^\mu n_b^\nu\,,
\ee
and where 
\be
I_{ab}\equiv S_{\mu\nu}n^\mu_a n^\nu_b
\ee
is the identity matrix of the screen space. Note again that ${\cal W}_{ab}$, as well as ${\cal R}$ and 
${\cal R}_{ab}$, are evaluated on the central geodesic and thus ${\cal W}_{ab}\left[x^\mu(\gr{n}^o,\hat v)\right]={\cal W}_{ab}(\gr{n}^o,\hat v)$. In terms of the electric and magnetic parts of the Weyl tensor, given respectively by \cite{Ellis:1998ct}
\be
{\cal E}_{\mu\nu}\equiv C_{\mu\rho\nu\sigma} u^\rho u^\sigma\,,\quad
{\cal B}_{\mu\nu}\equiv \frac{1}{2}\varepsilon_{\mu \alpha \beta \sigma}u^\sigma  
C_{\nu\rho}^{\phantom{\mu}\phantom{\rho}\alpha \beta} u^\rho\,,
\ee
the projected tensor ${\cal W}_{ab}$ becomes
\begin{equation}\label{proj_W}
{\cal W}_{ab}(\gr{n}^o,\hat v)= -2 n_{\langle a}^\mu n_{b\rangle}^\nu \left[{\cal E}_{\mu\nu} +
 {\cal B}_\mu^{\phantom{\nu}\sigma} \epsilon_{\sigma \nu}(\gr{n}) \right]_{\tiny\left\vert
\begin{array}{l}x^\mu(\gr{n}^o,\hat v)\\ \gr{n}(\gr{n}^o,\hat v)\end{array}\right.}
\end{equation}
In the expression above, $\langle\rangle$ stands for the traceless part with respect to $I^{ab}$. 
$\epsilon_{\mu\nu}(\gr{n})$ is the antisymmetric tensor in the projected space and is defined as
\begin{equation}
  \epsilon_{\mu\nu}(\gr{n}) \equiv u^\beta \varepsilon_{\beta \mu\nu \alpha } n^\alpha\,.
\end{equation} 
Now, ${\cal W}_{ab}$ being a spin-$2$ field, it can be decomposed in the helicity basis~(\ref{e:helicity}) as
\begin{equation} \label{e:Rpm}
{\cal W}_{ab}(\gr{n}^o,\hat v) \equiv - 2 \sum_{ \lambda=\pm}{\cal W}^\lambda (\gr{n}^o,\hat v) {n}^{\lambda}_a {n}^{\lambda}_b.
\end{equation}
This decomposition emphasizes once more that the two components ${\cal W}^\lambda$ are functions of $(\gr{n}^o,\hat v)$ alone, because they are evaluated on the lightcone. Recall that the $n_a^\lambda$ are constant so that we can use either $\gr{n}$ or $\gr{n}^o$ in Eq.~(\ref{e:Rpm}).

We now decompose the Jacobi map in terms of a convergence $\kappa$, a twist $V$ and a traceless shear $\gamma_{ab}$ as
\begin{equation}
 {\cal D}_{ab} \equiv \kappa I_{ab} + V \epsilon_{ab} + \gamma_{ab}\,,
\end{equation}
where
\[
\epsilon_{ab}= 2\ii n^-_{[a}n^+_{b]}\,.
\]
All these quantities are defined on our past lightcone so that we can also think of them as functions of $(\gr{n}^o,\hat v)$. The shear, being also a spin-2 field, is naturally decomposed similarly as
\begin{equation} \label{e:gammapm}
  {\gamma}_{ab}(\gr{n}^o,\hat v) \equiv \sum_{\lambda=\pm}{\gamma}^\lambda (\gr{n}^o,\hat v) {n}^{\lambda}_a {n}^{\lambda}_b\,.
\end{equation}
Finally, by inserting the decompositions~(\ref{e:Rpm}-\ref{e:gammapm}) in the Sachs equation~(\ref{gde2}) we find the desired equation of evolution
\begin{eqnarray}\label{e.evo1}
 \left(\!\frac{\dd^2}{\dd \hat v^2}+H_\parallel\frac{\dd}{\dd\hat v} - {\cal R}\!\right)
\left(\!\!\begin{array}{c}\kappa\\ \ii V\\ \gamma^\pm\end{array}\!\!\right)=
 -2 \left(\!\!\begin{array}{c}
  {\cal W}^{(-}\gamma^{+)} \\
  {\cal W}^{[-}\gamma^{+]}\\
  {\cal W}^\pm(\kappa\pm\ii V)
  \end{array}\!\!\right).
\end{eqnarray}
Note that, in practice, the integration of this system requires the evaluation of the past lightcone structure in order to determine $\gr{n}_i(\gr{n}^o,\hat v)$ and then $H_\parallel(\gr{n}^o,\hat v)$, ${\cal R}(\gr{n}^o,\hat v)$ and ${\cal W}^\pm(\gr{n}^o,\hat v)$.

\subsection{Multipole expansion} 
\label{multipole}
Equation (\ref{e.evo1}) is composed of scalars ($\kappa$, $V$, ${\cal R}$ and $H_\parallel$) and spin-$2$ fields 
($\gamma^\pm$ and ${\cal W}^\pm$) defined on the sphere. The former can be naturally decomposed in a basis of spherical harmonics as
\bea
\kappa(\gr{n}^o,\hat{v})&=&\sum_{\ell,m}\kappa_{\ell
  m}(\hat{v})Y_{\ell m}(\gr{n}^o)\\
V(\gr{n}^o,\hat{v})&=&\sum_{\ell,m}V_{\ell m}(\hat{v})Y_{\ell
  m}(\gr{n}^o)\\
{\cal R}(\gr{n}^o,\hat{v})&=&\sum_{\ell,m}{\cal R}_{\ell
  m}(\hat{v})Y_{\ell m}(\gr{n}^o)\\
H_\parallel(\gr{n}^o,\hat{v})&=&\sum_{\ell,m}h_{\ell m}(\hat{v})Y_{\ell m}(\gr{n}^o)
\eea
The latter, being spin-$2$ fields on the sphere, can be expanded on a basis of spin-weighted spherical harmonics~\cite{Goldberg1967} as 
\bea\label{EandBfromX}
  {\cal W}^\pm (\gr{n}^o,\hat v) &=& \sum_{\ell,m} \left[{\cal
      E}_{\ell m}(\hat v)\pm \ii {\cal B}_{\ell m}(\hat v)
  \right]Y_{\ell m}^{\pm 2}(\gr{n}^o)\,,\\
  {\gamma}^\pm (\gr{n}^o,\hat v) &=& \sum_{\ell,m} \left[{E}_{\ell m}(\hat v)\pm \ii {B}_{\ell m}(\hat v) \right]Y_{\ell m}^{\pm 2}(\gr{n}^o).
\eea
Note that $E$-modes are those having parity $(-1)^\ell$ while $B$-modes have parity $(-1)^{\ell+1}$~\cite{Pontzen2007}.

It is important to keep in mind that we are adopting an observer-based point of view so that all quantities are expressed in terms of $(\gr{n}^o,\hat v)$. In general, $\gr{n}(\gr{n}^o,\hat v)\not=\gr{n}^o$, with the obvious exception of e.g. FL spacetimes or for an observer at the center of symmetry of a Lema\^{\i}tre-Tolman spacetime. Part of the difficulty is thus contained in the determination of these coefficients, which include projection effects from the geodesic structure.

When inserting these decompositions in Eq.~(\ref{e.evo1}), products of spherical harmonics will appear on the r.h.s. They can be simplified using standard relations between spin-weighted spherical harmonics (see Appendix~\ref{AppA}). It follows that, in terms of multipoles, the equations of evolution for the convergence, twist and shear take the following general form
\begin{widetext}
{\small
\bea
  \frac{\dd^2 E_{\ell m} }{\dd \hat v^2} &=& \,{}^{2}C^{m m_1 m_2}_{\ell \ell_1 \ell_2} 
\left[\left({\cal R}_{\ell_1 m_1}-h_{\ell_1 m_1}\frac{\dd}{\dd\hat v}\right) 
\left( \delta_L^{+}E_{\ell_2 m_2} +\ii\delta_L^{-}B_{\ell_2 m_2} \right)
-2\kappa_{\ell_1 m_1}\left(\delta_L^{+}{\cal E}_{\ell_2m_2}+\ii\delta_L^{-}{\cal B}_{\ell_2m_2}\right)\right.\nonumber\\
&&\left.+2V_{\ell_1m_1}\left(- \ii \delta_L^{-}{\cal E}_{\ell_2m_2}+\delta_L^{+}{\cal B}_{\ell_2m_2}\right)
\right]\label{mastermultipoles1}\\
   \frac{\dd^2 B_{\ell m} }{\dd \hat v^2} &=& \,{}^{2}C^{m m_1 m_2}_{\ell \ell_1 \ell_2} \left[\left({\cal R}_{\ell_1 m_1}-h_{\ell_1 m_1}\frac{\dd}{\dd\hat v}\right)\left( \delta_L^{+}B_{\ell_2 m_2} -\ii\delta_L^{-}E_{\ell_2 m_2} \right)-2{\kappa}_{\ell_1 m_1} \left( \delta_L^{+}{\cal B}_{\ell_2 m_2} -\ii\delta_L^{-} {\cal E}_{\ell_2 m_2} \right) \right.\nonumber\\
&&\left.-2V_{\ell_1 m_1} \left( \delta_L^{-}\ii {\cal B}_{\ell_2 m_2}
    +\delta_L^{+} {\cal E}_{\ell_2 m_2} \right)
\right]\label{mastermultipoles2}\\
  \frac{\dd^2 \kappa_{\ell m} }{\dd \hat v^2} &=& \left\{\,{}^{0}C^{m m_1 m_2}_{\ell \ell_1 \ell_2}  \left( {\cal R}_{\ell_1 m_1}  \kappa_{\ell_2 m_2} - h_{\ell_1 m_1}  \frac{\dd \kappa_{\ell_2 m_2}}{\dd\hat v} \right)\right.\nonumber\\
&& \left.{-2(-1)^{m_1} {}^{2}C^{-m_2 -m
      m_1}_{\ell_2\;\ell\;\ell_1}}\left[ \delta_L^{+}({E}_{\ell_1
      m_1}{\cal E}_{\ell_2 m_2}+{B}_{\ell_1 m_1}{\cal B}_{\ell_2 m_2})
    +\ii\delta_L^{-} ({B}_{\ell_1 m_1}{\cal E}_{\ell_2
      m_2}-{E}_{\ell_1 m_1}{\cal B}_{\ell_2
      m_2})\right]\right\}\label{mastermultipoles3}
\eea
\bea
\frac{\dd^2 V_{\ell m} }{\dd \hat v^2} &=& \left\{\,{}^{0}C^{m m_1 m_2}_{\ell \ell_1 \ell_2}
\left( {\cal R}_{\ell_1 m_1}  V_{\ell_2 m_2} - h_{\ell_1 m_1}  \frac{\dd V_{\ell_2 m_2}}{\dd\hat v} \right)\right.\nonumber\\
&& \left.+2(-1)^{m_1} {}^{2}C^{-m_2 -m m_1}_{\ell_2\;\ell\;\ell_1}\left[\delta_L^{-}\ii({E}_{\ell_1 m_1}{\cal E}_{\ell_2 m_2}+{B}_{\ell_1 m_1}{\cal B}_{\ell_2 m_2}) -\delta_L^{+} ({B}_{\ell_1 m_1}{\cal E}_{\ell_2 m_2}-{E}_{\ell_1 m_1}{\cal B}_{\ell_2 m_2})\right]\right\}\label{mastermultipoles4}
\eea}
\end{widetext}
where 
\be
\delta_L^{\pm}\equiv [1\pm(-1)^L]/2\,,\quad L=\ell+\ell_1+\ell_2
\ee
and an implied sum over $\ell_1$, $\ell_2$, $m_1$, and $m_2$ is understood. This multipolar hierarchy for weak lensing, which does not rely on a particular background spacetime -- and on any perturbative expansion -- has never been derived before and sets the basis for general studies of the constraints on anisotropy and inhomogeneity from the weak-lensing $B$-modes. 

As soon as the spacetime has a non-vanishing Weyl tensor, $E$- and
$B$-modes are generated due to the coupling of the Weyl tensor to the
convergence and twist. It shares some similarities with the Boltzmann
hierarchy for the cosmic microwave background (see
e.g. Refs.~\cite{Pontzen2007,Hu2000}) but one needs to keep in mind
that ${\cal R}_{\ell m}$, $h_{\ell m}$, ${\cal E}_{\ell m}$, ${\cal
  B}_{\ell m}$ are non-local quantities since they have to be
evaluated on the geodesic.


\section{Applications to spatially homogeneous universes}

\subsection{Standard FL case}\label{FL}

In order to illustrate the formalism we consider the standard case of
(flat) FL spacetime with linear perturbations. At the background
level, the metric of the FL spacetime takes the simple form
\begin{equation}
 \dd s^2 =-\dd t^2 + a^2(t)\delta_{IJ}\dd x^I \dd x^J
\end{equation}
This spacetime enjoys 3 translational Killing vectors  $\lbrace\gr{e}_I\rbrace_{I\in\lbrace
  x,y,z\rbrace}\equiv \lbrace\frac{\partial}{\partial x^I}\rbrace_{I\in\lbrace
  x,y,z\rbrace}$ which define {\em everywhere}
a natural Cartesian basis. By normalizing these vectors, we can then define a triad of vectors
$\gr{e}_i$ whose components are $e_i^{\,I}=\delta_i^I/a$ (and their
associated 1-forms $\gr{e}^i$ whose components are
$e^i_{\,I}=\delta^i_{I} a$), that is a set of three orthonormal
space-like vectors (and forms) that can be used as a global Euclidian
basis. The set of vectors $\gr{n}_i^o$, which was a priori only
defined at the observer's position can then be defined everywhere by
imposing that their components $n_i^{o\,j}\equiv \gr{n}_i^o.\gr{e}^j$
in this reference basis remain the same everywhere. This enables to
compare ${\bm n}_i (\gr{n}^o,\hat v)$ to ${\bm n}^o_i$ even though
these sets of vectors are defined first at two different points of spacetime, as illustrated
in the right part of Fig.~\ref{f1}.

At the background level, the Weyl tensor vanishes (i.e. ${\cal
  E}_{\mu\nu}=0$ and ${\cal B}_{\mu\nu}=0$ are at least of order 1 in
perturbations) and the Ricci scalar, ${\cal R}^{(0)}$, depends only on
time. For this spacetime $\gr{n}(\gr{n}^o,\hat v)=\gr{n}^o$  for all
$\hat{v}$ so that the only nonzero multipolar coefficient
$h^{(0)}_{\ell m}$ is the monopole 
\be
  h^{(0)}_{00}=H\equiv \frac{\dot a}{a}\,
\ee 
where the dot refers to derivative with respect to $t$. From the expression above and the fact that 
$\dd\hat{v}=-\dd t$, we find from Eq.~(\ref{eqHparallel}) that $U\propto a^{-1}$. It then follows from Eq.~(\ref{def_z}) 
the well known result $1+z=a_0/a$. Moreover, since ${\cal E}^{(0)}_{\ell m}={\cal B}^{(0)}_{\ell m}=0$, it follows from Eqs. (\ref{mastermultipoles3}) and (\ref{mastermultipoles4}) that $\kappa^{(0)}_{00}$ and $V^{(0)}_{00}$ satisfy the same second order homogeneous equation of the form
\be
\frac{\dd^2 X^{(0)}_{00} }{\dd \hat v^2} = {\cal R}_{00}  X^{(0)}_{00} - H \frac{\dd X^{(0)}_{00}}{\dd\hat v}\,.
\ee
where $X^{(0)}_{00}$ stands for either $\kappa^{(0)}_{00}$ or $V^{(0)}_{00}$. The initial conditions~(\ref{initialconditions}) then lead to a homogeneous $\kappa^{(0)}_{00}$, given by the usual angular distance, and a vanishing twist so that\be
  \kappa^{(0)}_{00}=D_A\,, \qquad V^{(0)}_{00}=0\,.
\ee
Then, one concludes that
\be
E_{\ell m}^{(0)}=B_{\ell m}^{(0)}=0.
\ee\\

At first order in the perturbations, the perturbed metric with only
scalar perturbation reads in the Newton gauge
\be
 \dd s^2 =-(1+2 \Phi)\dd t^2 + a^2(t)(1-2\Psi)\delta_{IJ}\dd x^I \dd x^J\,,
\ee
where $\Phi$ and $\Psi$ are the two Bardeen potentials. The projected Ricci tensor is of the form \cite{Lewis2006}
\be
{\cal R}^{(1)}_{ab}=-D_{a}D_b(\Phi+\Psi)
\ee
where $D_a$ is the covariant derivative on the 2-sphere. It follows that 
\be
 {\cal B}^{(1)}_{ab}=0.
\ee
In the Born approximation (i.e. $\gr{n}(\gr{n}^o,\hat v)=\gr{n}^o$), only $h^{(0)}_{00}\not =0$ so that the r.h.s of Eqs.~(\ref{mastermultipoles1}-\ref{mastermultipoles2}) involves only $\,{}^{2}C^{m 0 m_2}_{\ell 0 \ell_2}$. Thus $\ell_2=\ell$ and $L$ is even (i.e. $\delta_L^-=0$). As a conclusion, in Eq.~(\ref{mastermultipoles1}) for the propagation of the $E$-modes, the only remaining term on the r.h.s. is 
\be
\left ({\cal R}^{(0)}_{00}-h^{(0)}_{00}\frac{\dd}{\dd\hat v}\right)E_{\ell m}^{(1)}-2 \kappa^{(0)}_{00}{\cal E}_{\ell m}^{(1)}
\ee
while in Eq.~(\ref{mastermultipoles2}) for the propagation of the $B$-modes it is 
\be
 \left({\cal R}^{(0)}_{00}-h^{(0)}_{00}\frac{\dd}{\dd\hat v}\right)B_{\ell m}^{(1)}.
\ee 
So we see that only $E$-modes are sourced, while $B$-modes would need to be initially non-zero to be non-vanishing today,
\begin{equation}
E_{\ell m}^{(1)}\not=0\, \qquad B_{\ell m}^{(1)}=0\,.
\end{equation}
Indeed, first order vector and tensor modes would generate $B$-modes since then ${\cal B}^{(1)}_{ab}\not=0$. 

The equation~(\ref{mastermultipoles3}) for the convergence has r.h.s. 
\be
\left({\cal R}^{(0)}_{00}-h^{(0)}_{00}\frac{\dd}{\dd\hat v}\right)\kappa_{\ell m}^{(1)}+{\cal R}_{\ell m}^{(1)}\kappa^{(0)}_{00},
\ee
as usual,\footnote{The ``standard'' convergence and shear are $-\kappa^{(1)}/\kappa^{(0)}$ and $\gamma_{ab}^{(1)}/\kappa^{(0)}$ in our notations.} since the other terms in Eq.~(\ref{mastermultipoles3}) are at least of second order in the perturbations. For the twist, the argument is similar but ${\cal R}_{\ell m}^{(1)}V^{(0)}_{00}=0$ and the initial conditions~(\ref{initialconditions}) imply $V^{(1)}_{\ell m}=0$. So, in conclusion
\begin{equation}
\kappa_{\ell m}^{(1)}\not=0\, \qquad   V_{\ell m}^{(1)}=0\,.
\end{equation}

At higher order, ${\cal E}_{ab}$ and ${\cal B}_{ab}$ are non-vanishing (note that one cannot simply drop out ${\cal B}_{\ell m}$ in the hierarchy, even for pure scalar modes), which leads to $B$-modes as well as twist. Moreover, projection effects and couplings induced by $h_{\ell m}$ need to be included; see Ref.~\cite{ordre2} for the case of second-order perturbations.

The absence of $B$-modes at first order in perturbations are due to the fact that
\begin{enumerate}
\item ${\cal B}^{(1)}_{\ell m}=0$ for scalar modes, 
\item at this order we can work in the Born approximation. 
\end{enumerate}
This latter point is extremely important since otherwise even if ${\cal B}_{\mu\nu}=0$ the dependence $\gr{n}(\gr{n}^o,\hat v)$ would generate a non-vanishing ${\cal B}_{ab}$~\cite{cooray}. Indeed, in Eqs.~(\ref{mastermultipoles1}-\ref{mastermultipoles3}) part of the difficulty lies in the determination of the coefficients ${\cal R}_{\ell m}$, $h_{\ell m}$, ${\cal E}_{\ell m}$ and ${\cal B}_{\ell m}$ that depend on the whole geodesic structure, as we shall now illustrate.

\subsection{Example of a Bianchi $I$} 
\label{BI}

We now consider the case of a spatially homogeneous but anisotropic universe described by a Bianchi $I$ spacetime for which the metric takes the form
\begin{equation}
 \dd s^2 =-\dd t^2 + a^2(t)\gamma_{IJ}(t)\dd x^I \dd x^J
\end{equation}
where the coordinates have been chosen so as to diagonalize
$\gamma_{IJ}(t)$. This solution is spatially homogeneous and the spatial shear
\begin{equation}
  \sigma_{IJ}\equiv \frac12\frac{\dd{\gamma}_{IJ}}{\dd t}
\end{equation} 
characterizes the spatial anisotropy, $a^2(t)\gamma_{IJ}$ being the
spatial metric, $a(t)$ is the volume averaged scale factor and
$\Theta=3H\equiv 3\dot a /a$ (see Refs.~\cite{PPU} for notations and properties). It follows that the kinematical quantities entering $H_\parallel$ in Eq.~(\ref{e.Hperp}) are
\be
 \Theta = 3H\,, \qquad
 \sigma _{IJ}\not=0\,,\qquad
 A_\mu=0\,.
\ee

Similarly to the FL case, this spacetime enjoys 3 Killing vectors
$\lbrace\gr{e}_I\rbrace_{I\in\lbrace x,y,z\rbrace}\equiv \lbrace\frac{\partial}{\partial x^I}\rbrace_{I\in\lbrace
  x,y,z\rbrace} $ which define {\em everywhere} a natural Cartesian
basis. Normalizing these vectors, we can then also define a triad of vectors
$\gr{e}_i$ that can be used as a global Euclidian basis. And similarly
to what has been done in the FL case, the set of vectors $\gr{n}_i^o$
can then be defined everywhere by imposing that their components in
this reference basis $\gr{e}_i$ remain the same, in order to allow the
comparison of ${\bm n}_i (\gr{n}^o,\hat v)$ to ${\bm n}^o_i$.

However, contrary to the FL case, one has to consider ({\em i}) the non-vanishing background electric Weyl tensor, 
\begin{equation}
{\cal E}^{(0)}_{IJ} = H \sigma_{IJ} + \frac{1}{3} \sigma^2 \gamma_{IJ} -\sigma_{IK}\sigma^K_J
\end{equation} 
while the magnetic part is identically null,
\begin{equation}
{\cal B}_{IJ}^{(0)}=0\,,
\end{equation} 
and ({\em ii}) the fact that at background level
$\gr{n}_i\not=\gr{n}^o_i$ (unless in the particular case of geodesics
along one of the three proper axis), which induces projection effects
so that $h_{\ell m}^{(0)}\not=0$.

The triad $\gr{n}_i(\gr{n}^o,\hat v)$ is related to the reference triad $\gr{n}_i^o$ by a rotation defined by three Euler angles as 
\bea\label{Euler1}
\gr{n}_i(\gr{n}^o,\hat v)&=&R_i^{\,\,j}(\alpha,\beta,\gamma)\gr{n}^o_j\nonumber\,,\\
&=&J_{\gr{n}_3^o}(\gamma)_i^{\,\,j} J_{\gr{n}^o_2}(\beta)_j^{\,\,k} J_{\gr{n}_3^o}(\alpha)_k^{\,\,l} \,\gr{n}_l^o 
\eea
where  the Euler angles are also functions of $(\gr{n}^o,\hat
v)$. 
The determination of $\alpha$, $\beta$ and $\gamma$ requires the integration of the geodesic equation in the Bianchi spacetime. 

Then, for a typical tensor ${T}_{\mu\nu}$ at an event $x^\mu$, its projection
orthogonally to $\gr{n}$, i.e. its components
${T}^\pm[x^\mu,\gr{n}_i]\equiv {T}^\pm\left[x^\mu(\gr{n}^o,\hat v)\right]$ in
the helicity basis $\gr{n}_\pm$, can be related to its projection at
the same event $x^\mu$, but orthogonally to $\gr{n}^o$ with components
${T}_o^\pm[x^\mu,\gr{n}^o_i]$ in the helicity basis $\gr{n}^o_\pm$. For
a spin $s$ tensor, this transformation reads (see details in appendix~\ref{AppB}) in general
\begin{equation}\label{Transformationgen}
   {T}^\pm[x^\mu,\gr{n}_i] = \exp(\pm \ii s \phi) \exp(\beta^a
   D_a) {T}^\pm_o(x^\mu,\gr{n}^o_i), 
\end{equation}
with 
\[
\phi\equiv\alpha+\gamma\quad\mbox{and}\quad\bm{\beta}\equiv\beta[\gr{n}_1^o\cos\gamma+\gr{n}_2^o\sin\gamma]\,.
\]
For a homogeneous spacetime, the dependence in $x^\mu$ of ${T}^\pm_o$ reduces to a time
dependence. Eq.~(\ref{Transformationgen}) evaluated for a rank-$2$
tensor (that is $s=2$) is needed to account for the
projection effects in the definition of ${\cal
  W}^\pm[x^\mu(\gr{n}^o,\hat v)]$. Similarly, Eq.~(\ref{Transformationgen})
in the case $s=0$ (that is for a scalar field) is needed for the
projection effects of ${H}_\parallel[x^\mu(\gr{n}^o,\hat v)]$ and ${\cal
  R}[x^\mu(\gr{n}^o,\hat v)]$. 
The Weyl tensor having only a non-vanishing electric part (with only
non-vanishing components ${\cal E}_{xx}$, ${\cal E}_{yy}$ and ${\cal
  E}_{zz}$ in the natural Cartesian basis), one has
\begin{equation}\label{EBdebase}
   {\cal W}^{\pm}_o(\eta,\gr{n}_i^o) = \sum_{m=0,\pm2} {\cal E}^o_{2 m}[\eta(\hat{v})]Y_{2\, m}^{\pm 2}(\gr{n}_i^{o})
\end{equation}
with 
\bea
{\cal E}^o_{2\,0}&=&\sqrt{\frac{2 \pi}{15}}\left(2{\cal E}_{zz}-{\cal
    E}_{xx}-{\cal E}_{yy}\right)\\
{\cal E}^o_{2,\pm2}&=&\sqrt{\frac{\pi}{5}}({\cal E}_{xx} - {\cal E}_{yy})\,. 
\eea
The projection of  the electric Weyl tensor has a directional dependence for $\ell=2$ and $m=0,\pm2$. However, the directional dependence of $\phi$ and $\beta^a$ in Eq.~(\ref{Transformationgen}), i.e. the projection effects, sources and mixes $E$ and $B$ modes at higher $\ell$, as for CMB polarization $E/B$ modes mixing~\cite{Challinor2005}. This projection effect also induces non-vanishing ${\cal R}_{\ell m}$ terms even if the background Ricci is homogeneous.\\

To go further and understand how this mixing of $E$ and $B$ modes
arises, let us assume that $\hat\sigma^2/\Theta^2$ is small, so that
we can work at first order on this parameter (we can think of B$I$ has a homogeneous
perturbation of FL). Then, the geodesic equation and the parallel transport of ${\bm n}_a$ [Eq.~(\ref{e:prop_n})] 
lead to
\be
\frac{\dd}{\dd \hat v} n^i = S^{ik}\sigma_{kj}n^j\,,\qquad
S_{ij}\frac{\dd}{\dd \hat v} n_a^j=0
\ee
and thus at lowest order one easily obtains that $\phi\simeq 0$ and $\beta^a(\gr{n}^o,\hat{v}) \simeq
\int_0^{\hat{v}} D_a \sigma(\gr{n}^o,\hat{v}') \dd \hat{v}'$. Here $\sigma(\gr{n},\hat{v}) \equiv \sigma_{IJ}(\hat{v}) n^I n^J/2$ 
can be thought as a lensing potential and Eq. (\ref{Transformationgen})
for $\bm{\mathcal{W}}$ gives
\begin{equation}\label{lensBianchi}
{\cal W}^\pm[x^\mu,\gr{n}_i]\simeq [1+\beta^a D_a] {\cal W}^\pm_o(\eta,\gr{n}_i^o),
\end{equation}
similar to the form for linearized lensing in FL~\cite{Hu2000,Challinor2002} on
light polarization. $\sigma(\gr{n}^o,\hat{v})$ obviously contains only $\ell=2$ multipoles. Because of the derivative coupling, using \cite{Hu2000}
\be\label{LensingYlms}
D_a Y_{\ell_1 m_1} D^a Y_{\ell_2 m_2}^{\pm s}=\sum_{\ell m} L_{\ell \ell_1 \ell_2}\,{}^{\pm s}C^{m m_1 m_2}_{\ell \ell_1 \ell_2}\, Y_{\ell m}^{\pm s}
\ee
where
\be
L_{\ell \ell_1 \ell_2}\equiv\frac{1}{2}\left[\ell_1(\ell_1+1)+\ell_2(\ell_2+1)-\ell(\ell+1)\right]
\ee
and further defining 
$$
{}^{\pm s}I^{m m_1 m_2}_{\ell \ell_1 \ell_2}\equiv  L_{\ell \ell_1 \ell_2}\,{}^{\pm s}C^{m m_1 m_2}_{\ell \ell_1 \ell_2},
$$ 
one can convince oneself that, at background level, terms such as 
\be
{\cal E}^{(0)}_{\ell m} \simeq {\cal E}^{o}_{\ell m}+ \,{}^{2}I^{m m_1
  m_2}_{\ell \ell_1 \ell_2}  \left(\int_0^{\hat{v}} {\sigma}_{\ell_1 m_1} \dd \hat{v}'\right)\delta_L^{+}{\cal E}^o_{\ell_2 m_2} 
\ee
 and 
\be
{\cal B}^{(0)}_{\ell m} \simeq -\ii \,{}^{2}I^{m m_1 m_2}_{\ell \ell_1
  \ell_2} \left(\int_0^{\hat{v}} {\sigma}_{\ell_1 m_1} \dd \hat{v}' \right)\delta_L^{-}{\cal E}^o_{\ell_2 m_2} 
\ee
 are expected when extracting the $E$ and $B$ modes out of Eq.~\ref{lensBianchi} . Thus a multipolar $\ell = 4$ $B$-mode will appear. One needs however to rely on the full transformation~(\ref{Transformationgen}) so that the $E$- and $B$-modes shall be generated for larger $\ell$'s. Similar sources arise from ${\cal R}$ and $H_\parallel$ for which projection effects will generate non-vanishing ${\cal R}_{\ell m}^{(0)}$ and $h_{\ell m}^{(0)}$.

A full analysis, including perturbations and magnitude estimations will be presented in Ref.~\cite{next}. Our argument sketches the expected effects that arise from the higher multipoles induced by the background Weyl tensor and the fact that $\gr{n}\not=\gr{n}^o$, an effect that cannot be neglected even in the Born approximation for anisotropic spaces. Besides, in B$I$ spacetimes, the amplitude of vectors and tensors is of order of the shear times the amplitude of the scalars, another source of $B$-modes.

\section{Conclusion} 
\label{conclusion}

We have provided a new multipolar hierarchy for weak lensing. Our formalism, which is fully covariant, does not rely on perturbation theory nor on the choice of a background spacetime. It allows us to relate the property of the shear to symmetry properties of the background spacetime and discuss the generation of $B$-modes. We have argued that a violation of local isotropy is expected to leave a $B$-mode signature on all scales. This result is important for future surveys, such as the Euclid mission~\cite{euclid} (early results on the $B$-modes have already been obtained from CFHTLS~\cite{cfhtls} and DLS~\cite{lsstpaper} and we can forecast that Euclid will typically decrease the error bars on the $B$-modes by a factor of order 10-40 on scales ranging up to 40 degree, that is in the linear regime where astrophysical sources of $B$-modes are expected to be negligible) and may us allow to set new constraints on the deviation from spatial isotropy on cosmological scales. The quantitative computation of the level of $B$-modes expected on large scales, where the gravitational dynamics can be considered linear, for a Bianchi universe is currently being investigated \cite{next} and requires to study in details the cosmological perturbation theory beyond the analysis of a scalar field \cite{PPU}.

\acknowledgements 
We thank Yannick Mellier and Francis Bernardeau for their comments and insights and Anthony Tyson for bringing the reference \cite{lsstpaper} to our attention. TSP thanks the {\it Institut d'Astrophysique de Paris} for the support and hospitality during the early stages of this work.

\appendix
\section{Spin-weighted spherical harmonics}\label{AppA}
We gather here a few important identities and relations between spin-weighted spherical harmonics used in this text. The reader is referred to Ref.~\cite{Goldberg1967} for more details.

Spin-weighted spherical harmonics form a complete set of orthonormal functions on the sphere, satisfying
\be
\int\dd^2\gr{n}\, Y^{\star\,\pm s}(\gr{n})_{\ell m}Y^{\pm s}_{\ell'm'}(\gr{n})=\delta_{\ell\ell'}\delta_{mm'}\,.
\ee
An important identity, used in particular to derive Eqs. (\ref{mastermultipoles1})-(\ref{mastermultipoles4}), is
\be
Y_{\ell_1 m_1} Y_{\ell_2 m_2}^{\pm s}=\sum_{\ell m} \, {}^{\pm s}C^{m m_1 m_2}_{\ell \ell_1 \ell_2} \,Y_{\ell m}^{\pm s}
\ee
where
\be
{}^{\pm s}C^{m m_1 m_2}_{\ell \ell_1 \ell_2}\equiv\int \dd^2 \gr{n} Y_{\ell m}^{\pm s \star}(\gr{n}) Y_{\ell_1 m_1}(\gr{n}) Y_{\ell_2 m_2}^{\pm s}(\gr{n})\,.
\ee
Using the transformation of spherical harmonics under parity, it can be shown that the above coefficients satisfy
\be
{}^{\mp s}C^{m m_1 m_2}_{\ell \ell_1 \ell_2}=(-1)^{L} \,\,{}^{\pm s}C^{m m_1 m_2}_{\ell \ell_1 \ell_2}
\ee
where $L=\ell+\ell_1+\ell_2$.

\section{Expansion of tensors on the sphere}
\label{AppB}

The transformation~(\ref{Euler1}) of the triad is easily computed for the tangent space
basis when the helicity vectors are used. Indeed we first note that for a general rotation
around an axis $\gr{n}$, the corresponding helicity basis at the point
of the sphere, that is $\gr{n}_\pm(\gr{n})$, transforms as
\begin{equation}\label{Magichelicity}
J_{\gr{n}}(\varphi) \cdot \gr{n}_\pm = \exp^{\mp \ii \varphi} \gr{n}_\pm\,,
\end{equation}
i.e., it is a diagonal matrix in this basis.
Furthermore, when the axis of the rotation does not coincide
with the direction defining the helicity basis, the rotation is still diagonal in the
sense that, if we consider a general rotation $\gr{J}$ not necessarily around
the axis $\gr{n}$, then  
\be \label{RotationDet1}
\gr{n}_\mp^\star(\gr{J}\cdot\gr{n})\cdot\gr{J}\cdot\gr{n}_\pm(\gr{n})=0
\ee
or $\gr{J} \cdot \gr{n}_\pm(\gr{n}) \propto \gr{n}_\pm(\gr{J}\cdot\gr{n})$, where we recall that 
$\gr{n}_\pm(\gr{J}\cdot\gr{n})$ is the helicity basis at the point of the sphere which is the image of
$\gr{n}$ by $\gr{J}$. It comes essentially from the fact that rotations defined as $SO(3)$ are not only keeping orthogonality conditions, but also orientations of triads as their determinant is required to be $1$. 
 
In order to use these interesting properties, we reformulate the transformation~(\ref{Euler1}), which is
written with rotations around the axis of a fixed frame, by its form
where the rotations are performed around the axis of the rotating
frame. In that case it reads
\be\label{Euler2}
\gr{n}_i(\gr{n}^o,\hat{v})=J_{\gr{n}_3^{''o}}(\alpha)_i^{\,\,j} J_{\gr{n}^{'o}_2}(\beta)_j^{\,\,k} 
J_{\gr{n}_3^o}(\gamma)_k^{\,\,l} \,\gr{n}_l^o 
\ee
with $\gr{n}^{'o}_2 \equiv J_{\gr{n}_3^o}(\gamma)\cdot\gr{n}_2^o$ and
$\gr{n}_3^{''o} \equiv J_{\gr{n}^{'o}_2}(\beta)\cdot\gr{n}_3^o$. Using
the transformation rule~(\ref{Magichelicity}) and the property~(\ref{RotationDet1}), we deduce that the
transformation rule for the helicity vectors is just
\begin{equation}\label{Eulersimple}
\gr{n}_\pm(\gr{n}^o,\hat{v})=\exp^{\mp \ii (\alpha+\gamma)}J_{\gr{n}^{'o}_2}(\beta)\cdot\gr{n}_\pm^o \,.
\end{equation}

It then proves convenient to express a rotation as a parallel
transport. Indeed under any rotation of angle $\varphi$ around an axis
$\gr{n}_{\rm rot}$, a tensorial quantity $\gr{T}$ in the tangent space at a point
$\gr{n}_{\rm equator}$ of the corresponding equator (that is such that
$\gr{n}_{\rm equator}\cdot\gr{n}_{\rm rot}=0$) is transformed exactly
as if it were parallel transported with the vector $\gr{n}_{\rm transport}\equiv\varphi \gr{n}_{\rm
rot} \times \gr{n}_{\rm equator}$. Let us note this parallel transport ${\cal T}_{\gr{n}_{\rm transport}} (\gr{T})$.
We insist that this rephrasing of a rotation as a parallel transport is valid only on the equator of the
rotation. In our case, this is enough to reformulate our transformation~(\ref{Eulersimple}) as
\begin{equation}\label{Eulersupersimple}
\gr{n}_\pm(\gr{n}^o,v)=\exp^{\mp \ii (\alpha+\gamma)} {\cal T}_{{\bm \beta}} (\gr{n}_\pm^o) 
\end{equation}
with
\be
{\bm \beta}(\gr{n}^o)\equiv \beta[\gr{n}^o_1\cos\gamma+\gr{n}^o_2\sin\gamma]\,
\ee
and ${\cal T}_{{\bm \beta}}$ being the parallel transport along ${\bm \beta}$.

Let us apply the result~(\ref{Eulersupersimple}) to obtain an expression for the expansion of 
a tensor on the sphere. For a rank-$s$ tensor $\gr{T}$, its projection
orthogonally to $\gr{n}$ defines a tensor field on the sphere. For
instance, the rank-$2$ tensor ${\cal W}_{\mu\nu}$ defines a tensor field on
the sphere $S_\mu^\rho S_\nu^\sigma {\cal W}_{\rho\sigma}$ given that the
screen projector ${S}_{\mu\nu}$ depends on the position $\gr{n}$  on the sphere
of directions. In the evaluation of
the geodesic deviation equation, we are led to express the
components $T^\pm \equiv T^\pm[\gr{n}(\gr{n}^o,\hat v)]$ of a 
tensor field at a point $\gr{n}(\gr{n}^o,\hat v)$ in the helicity basis
$\gr{n}_\pm(\gr{n}^o,\hat v)$ in function of its components $T_o^\pm \equiv
T_o^\pm(\gr{n}^o)$  at a reference point $\gr{n}^o$ in the
helicity basis $\gr{n}_\pm^o$. This expansion is obtained as follows
\bea
T^\pm& \equiv& \gr{T}\cdot\gr{n}_\mp \dots
\gr{n}_\mp|_{\gr{n}(\gr{n}^o,\hat v)}\\
&=&{\cal T}^{-1}_{\bm \beta}(\gr{T}\cdot\gr{n}_\mp \dots
\gr{n}_\mp)|_{\gr{n}^o}\nonumber\\
&=&\exp^{\pm \ii s(\alpha+\gamma)} {\cal T}^{-1}_{\bm
  \beta}(\gr{T})|_{\gr{n}^o}\cdot\gr{n}^o_\mp \dots
\gr{n}^o_\mp\nonumber \\
&=&\exp^{\pm  \ii s(\alpha+\gamma)} [\exp{(\beta^a D_a)}\gr{T}|_{\gr{n}^o} ]\cdot\gr{n}^o_\mp \dots
\gr{n}^o_\mp\nonumber
\eea
From the first to the second line, we have used that a scalar field
(the components of $\gr{T}$) evaluated
in $\gr{n}$ or its parallel transport back along ${\bm \beta}$
evaluated in $\gr{n}_o$ are equal. From the second to the third line, we
have used the transformation rule~(\ref{Eulersupersimple}) of the
triad. From the third to the fourth line, we have used the
exponentiation of the parallel transport in terms of covariant
derivatives $D_a$ on the 2-sphere. Then, with a common abuse of
notation (see for instance the discussion at the end of Ref.~\cite{Challinor2002}), which we also consistently use in
Eq.~(\ref{LensingYlms}), this is rewritten in a short form as
\be
T^\pm\equiv \exp^{\pm \ii s(\alpha+\gamma)} \exp{(\beta^a D_a)}T_o^\pm \,.
\ee


\end{document}